\documentclass[aps,showpacs]{revtex4}
\usepackage{graphicx}
\begin{document}

\title{Theory of collective Raman scattering from a Bose-Einstein condensate}
\author{Mary M. Cola and Nicola Piovella}
\affiliation{Dipartimento di Fisica, Universit\`a degli Studi di
Milano \& INFM, Via Celoria 16, Milano I-20133, Italy.}

\begin{abstract}
Recent experiments have demonstrated superradiant Raman scattering
from a Bose-Einstein condensate driven by a single off-resonant
laser beam. We present a quantum theory describing this
phenomenon, showing Raman amplification of matter wave due to
collective atomic recoil from 3-level atoms in a
$\Lambda$-configuration. When atoms are initially in a single lower internal state,
a closed two-level system is realized between atoms with
different internal states, and entangled atom-photon pairs can be generated.
When atoms are initially prepared in both the lower internal states, a fraction of atoms
recoiling in the backward direction can be generated.
\end{abstract}

\maketitle

Important progress in the study of the coherent interaction
between atoms and photons have been recently obtained using
Bose-Einstein condensates (BEC) of low-density alkali atoms
\cite{BEC}. In the case where the atoms interact only with far
off-resonant optical fields, the dominant atom-photon interaction
is two-photon Rayleigh scattering. In this situation, collective atomic
recoil lasing (CARL) \cite{CARL,moore:1,moore:2,gatelli} causes exponential
enhancement of the number of scattered photons and atoms.
Experimentally, CARL from a BEC has been observed
so far in the Superradiant regime \cite{MIT:1,kozuma,MIT:2,LENS}, in
which photons are scattered into the end-fire modes along the
major axis of an elongated condensate. In these experiments, the
atoms after the collective scattering remain in the original
internal, gaining a recoil momentum
$\hbar({\vec k_2}-{\vec k_1})$, where ${\vec k_2}$ and ${\vec k_1}$
are the wave vectors of the pump and scattered photons,
respectively. The scattered atoms may experience further
collective scattering, leading to the observed superradiant
cascade \cite{MIT:1}.

In two recent experiments \cite{Raman:1,Raman:2} it has been observed
superradiant Raman scattering, in which the atoms remain, after the process,
in a different hyperfine state not resonant with the pump
laser beam. As a consequence, no further scattering of pump
photons occurs. In this Brief Report, we present a theory of the collective atomic recoil
lasing from a 3-level atomic BEC which describes the observed phenomena. In particular,
the theory demonstrates that maximum atom-photon entanglement can be generated in this system.

We consider a cloud of BEC atoms which have three internal states
labeled by $|b\rangle$, $|c\rangle$, and  $|e\rangle$ with
energies $E_b<E_c<E_e$, respectively. The two lower
states   $|b\rangle$ and  $|c\rangle$ can be hyperfine states in
each of which the atoms can live for a long time. They are
coupled to the  upper state  $|e\rangle$ via, respectively, a
classical pump field and a quantized probe field
of frequencies $\omega_2$ and $\omega_1$  in the $\Lambda$-configuration.
The interaction scheme is shown in Fig. 1.

The second quantized Hamiltonian to describe the system at zero
temperature is given by
\begin{equation}
\label{ham}
\hat{H}=\hat{H}_{atom}+\hat{H}_{atom-field},
\end{equation}
where $\hat{H}_{atom}$ gives the free evolution of the the atomic
fields and $\hat{H}_{atom-field}$ describes the dipole
interactions between the atomic fields and the pump and probe fields.
We assume the condensate to be sufficiently diluite  in order to neglect the
atom-atom interaction.
The free atomic Hamiltonian is given by
\begin{equation}
\label{hamatom} \hat{H}_{atom}=\sum_{\alpha=b,c,e}\int d^3x
\hat{\psi}^{\dagger}_{\alpha}({\vec x},t) \left
[-\frac{\hbar^2}{2m}\nabla^2\right] \hat{\psi}_{\alpha}({\vec x},t),
\end{equation}
where $\hat{\psi}_{\alpha}({\vec x},t)$ and
$\hat{\psi}^{\dagger}_{\alpha}({\vec x},t)$ are the boson
annihilation  and creation operators in the interaction picture
for the $|\alpha\rangle$-state atoms at position ${\vec x}$,
respectively. They satisfy the standard boson commutation relation
$[\hat{\psi}_{\alpha}({\vec x},t),
\hat{\psi}^{\dagger}_{\beta}({\vec x}',t)] =\delta_{\alpha
\beta}\delta({\vec x}-{\vec x}')$ and $[\hat{\psi}_{\alpha}({\vec
x},t), \hat{\psi}_{\beta}({\vec x}',t)]
=[\hat{\psi}^{\dagger}_{\alpha}({\vec
x},t),\hat{\psi}^{\dagger}_{\beta}({\vec x}',t)]=0$.

The atom-laser interaction in the dipole approximation is
described by the following Hamiltonian
\begin{eqnarray}
\label{hamatomfield} \hat{H}_{atom-field} = -\hbar \int d^3x
\left[ \frac{1}{2}\Omega\hat{\psi}^{\dagger}_e({\vec x},t)
\hat{\psi}_b({\vec x},t)e^{i({\vec k_2}\cdot {\vec
x}-\Delta_2t)}\right.\nonumber \\
+\left.g_1\hat{a}_1(t)\hat{\psi}^{\dagger}_e({\vec
x},t)\hat{\psi}_c({\vec x},t) e^{i({\vec k_1}\cdot{\vec
x}-\Delta_1t)}+H.c.\right],
\end{eqnarray}
where $\omega_{b,c}=(E_e-E_{b,c})/\hbar$ are the resonant
frequencies for the two atomic transitions,
$\Delta_2=\omega_{2}-\omega_{b}$,
$\Delta_1=\omega_{1}-\omega_{c}$, $g_1=\mu_{ce}{\cal E}_1/\hbar$
and $\Omega=\mu_{be}E_2/\hbar$ with $\mu_{\alpha \beta}$ denoting
a transition dipole-matrix element between states $|\alpha\rangle$
and $|\beta\rangle$, ${\cal
E}_1=\sqrt{\hbar\omega_1/2\epsilon_0V}$ being the electric field
per photon for the quantized probe field of frequency $\omega_1$
in a mode volume $V$, and $E_2$ being the amplitude of the
electric field for the classical pump laser beam of frequency
$\omega_2$. Finally, $\hat{a}^{\dagger}_1(t)$ and
$\hat{a}_1(t)$ are photon creation and annihilation operators for
the probe field, satisfying the boson commutation relation
$[\hat{a}_1(t), \hat{a}^{\dagger}_1(t)]=1$.

We consider the case where the pump laser is detuned far enough
away from the atomic resonance that the excited state population
remains negligible, a condition which requires that
$\Delta_2\gg\gamma_e$, where $\gamma_e$ is the natural width of
the atomic transition between the excited state $|e\rangle$ and
the hyperfine ground state $|b\rangle$. In this regime the atomic
polarization adiabatically follows the ground state population,
allowing the formal elimination of the excited state atomic field
operator. Writing the Heisenberg equation for
$\hat\psi_e\exp[i({\vec k}_2\cdot{\vec x}-\Delta_2 t)]$ and dropping
the kinetic term, we obtain
\begin{equation}
\hat{\psi}_e({\vec x},t)\approx  -\frac{1}{\Delta_2}\left\{
\frac{1}{2}\Omega\hat{\psi}_b({\vec x},t)+
g_1\hat a_1(t)\hat{\psi}_c({\vec x},t) e^{-i\theta+i\delta
t}\right\} e^{i({\vec k}_{2}\cdot{\vec x}-\Delta_2 t)}
\label{adiabatic}
\end{equation}
where $\theta=({\vec k}_2-{\vec k}_1)\cdot{\vec x}$ and
$\delta=\Delta_2-\Delta_1=\omega_2-\omega_1-\Delta_{cb}$, with
$\Delta_{cb}=(E_c-E_b)/\hbar$. Substituting Eq.(\ref{adiabatic})
into Eq.(\ref{hamatomfield}) and neglecting the small light shifts
proportional to $|\Omega|^2$ and $g_1^2\hat a_1^\dag\hat a_1$, we
arrive at the following effective Hamiltonian:
\begin{equation}
\hat{H}=\sum_{\alpha=b,c}\int d^3x
\hat{\psi}^{\dagger}_{\alpha}({\vec x},t) \left
[-\frac{\hbar^2}{2m}\nabla^2\right] \hat{\psi}_{\alpha}({\vec x},t)
 +i\hbar g\int d^3x \left [\hat a^\dag\hat{\psi}_b({\vec
x},t)\hat{\psi}^{\dagger}_c({\vec x},t)e^{i\theta} -{\rm
H.c.}\right ] -\hbar\delta\hat a^\dag\hat a, \label{H_eff}
\end{equation}
where $g=g_1\Omega/2\Delta_2$ and $\hat a=i\hat a_1 e^{i\delta t}$.

Neglecting shape effects due to the finite size of the condensate, we can perform the expansion
on momentum eigenstates \cite{gatelli}:
\begin{equation}
\hat{\psi}_{b}={\cal C}\sum_{n=-\infty}^{+\infty}\hat{b}_{n}e^{in\theta} \qquad
\hat{\psi}_{c}={\cal C}\sum_{n=-\infty}^{+\infty}\hat{c}_{n}e^{in\theta}
\label{sum}
\end{equation}
where  $[\hat{c}_{n},\hat{c}_{m}^{\dag}]=\delta_{n,m}$,
$[\hat{b}_{n},\hat{b}_{m}^{\dag}]=\delta_{n,m}$,
$[\hat{b}_{n},\hat{c}_{m}]=[\hat{b}_{n},\hat{c}^{\dag}_{m}]=0$ and
${\cal C}$ is a normalization constant.
Substituting Eqs.(\ref{sum}) into Eq.(\ref{H_eff}), the Hamiltonian becomes:
\begin{equation}
\hat{H}=\sum_{n=-\infty}^{+\infty}\{\hbar\omega_{r}n^{2}
(\hat{b}^{\dag}_{n}\hat{b}_{n}+\hat{c}^{\dag}_{n}\hat{c}_{n})+
i\hbar g(\hat{a}^{\dag}\hat{c}^{\dag}_{n}\hat{b}_{n-1}-{\rm H.c.})\}
-\hbar \delta \hat{a}^{\dag}\hat{a}.
\label{ham2}
\end{equation}
and the Heisenberg equations for $\hat b_n$, $\hat c_n$ and $\hat a$ are
\begin{eqnarray}
\frac{d\hat{b}_{n}}{dt}&=&-i\omega_{r}n^2\hat{b}_{n}-g\hat{a}
\hat{c}_{n+1}\label{bn}\\
\frac{d\hat{c}_{n}}{dt}&=&-i\omega_{r}n^2\hat{c}_{n}+g\hat{a}^{\dag}
\hat{b}_{n-1}\label{cn}\\
\frac{d \hat{a}}{d t}&=&i\delta \hat{a}+ g\sum_{n}\hat{b}_{n}\hat{c}^{\dag}_{n+1}\label{bncn},
\end{eqnarray}
where $\omega_r=\hbar q^2/2m$ is the recoil frequency and
$\hbar{\vec q}=\hbar({\vec k}_2-{\vec k}_1)$ is the photon recoil
momentum. In Eqs.(\ref{sum}),  $\hat b_n$ and $\hat c_n$ are
annihilation operators for the modes $|b,n\rangle$ and
$|c,n\rangle$, corresponding to atoms in the internal state
$|b\rangle$ and $|c\rangle$, respectively, and with momentum ${\vec
p}=n\hbar{\vec q}$. Notice that Eqs.(\ref{bn})-(\ref{bncn})
conserve the total number of atoms $N$, i.e. $\sum_n\{\hat b_n^\dag\hat
b_n+\hat c_n^\dag\hat c_n\}=\hat N$, and the total momentum $\hat
Q=\hat a^\dag\hat a+\sum_n n\{\hat b_n^\dag\hat b_n+\hat
c_n^\dag\hat c_n\}$. Furthermore, the number of atoms in the subsystem
${\cal C}_n=\{|b,n\rangle,|c,n+1\rangle\}$ is also conserved, i.e. $\hat
b_n^\dag\hat b_n+\hat c_{n+1}^\dag\hat c_{n+1}=\hat N_n$ is a
constant for every $n$. This means that each subsystem ${\cal
C}_n=\{|b,n\rangle,|c,n+1\rangle\}$ is closed. However, atoms belonging to
different ${\cal C}_n$ are coupled by the common radiation field
$\hat a$.

The system of Eqs.(\ref{bn})-(\ref{bncn}) describes the two-photon Raman scattering,
in which an atom is transferred from the state $|b,n\rangle$ to the state
$|c,n+1\rangle$ when it scatters a photon from the pump to the
probe, i.e. when it "emits" a probe photon, whereas the atom is
transferred from the state $|c,n\rangle$ to the state
$|b,n-1\rangle$ when it scatters a photon from the probe to the
pump, i.e. when it "absorbs" a probe photon.
The main difference with respect to the normal CARL regime is that
after emission of a probe photon the atom changes the internal
state from $|b\rangle$ to $|c\rangle$. In particular, if
atoms are initially in the internal state $|b\rangle$, they can only
emit probe photons. As a consequence, in the Superradiant regime,
in which emission dominates over absorbtion, atoms are
transferred from the initial state $|b,0\rangle$ to the final
state $|c,1\rangle$, where they can not anymore emit probe photons,
experiencing subsequent superradiant scattering. Hence, when atoms
are initially in the state $|b,0\rangle$, the condensate
behaves as a closed two-level system.

In the linear regime where $N_{c1}\ll N_b$, where $N_b$ and $N_{c1}$ are the number
of atoms in the initial state $|b,0\rangle$ and in the recoiling state $|c,1\rangle$,
we may assume $\hat{b}_{0}\approx\sqrt{N_b}$ and the Hamiltonian
(\ref{ham2}) reduces to:
\begin{equation}
\hat{H}_{eff}=\hbar\omega_{r}\hat{c}^{\dag}_{1}\hat{c}_{1}+ i\hbar
g\sqrt{N_b}(\hat{a}^{\dag}\hat{c}^{\dag}_{1}-\hat{a}\hat{c}_{1})
-\hbar \delta \hat{a}^{\dag}\hat{a}. \label{ham3}
\end{equation}
This means that we are investigating a system which is
analogous to the non-degenerate optical parametric amplifier (OPA)
\cite{She84,WalMil94} and involves the generation of correlated
atom-photon pairs. The evolved state at time $t$ is a pure
bipartite state
\begin{equation}
|\psi\rangle = \frac{1}{\sqrt{1+\langle\hat n_c\rangle}} \sum_{n=0}^\infty
\left(\frac{\langle\hat n_a\rangle}{1+\langle\hat n_c\rangle}\right)^{n/2}
e^{in\phi}|n,n\rangle,\label{entanstate}
\end{equation}
where $\langle\hat n_a\rangle=\langle\hat a^\dag\hat a\rangle$ and
$\langle\hat n_c\rangle=\langle\hat c_1^\dag\hat c_1\rangle$. Eq.(\ref{entanstate})
shows maximal entanglement between atoms and photons
(according to the excess von Neumann entropy criterion \cite{pho})
and has the same form of the twin-beam state of radiation generated from an OPA and
used to realize continuous variable optical teleportation \cite{kim}. The idea of using Raman-scattering
from an optically driven BEC as a source of atom-photon pairs was originally proposed by
Moore and Meystre \cite{moore:3}, however without exploiting the amplification CARL process.
In the ordinary quantum CARL a detailed theory for the interaction
of quantized atomic and optical fields in the linear regime has
been developed, with emphasis on the manipulation and control of
their quantum statistics and the generation of quantum
correlations and entanglement between matter and light waves
\cite{moore:1,moore:2,mary} . From such model it results that, in
the linear regime, the quantum CARL Hamiltonian reduces to that for three
coupled modes, the first two modes corresponding to atoms having
lost or gained a quantum recoil momentum $\hbar\vec q$ in the
two-photon Bragg scattering between the pump and the probe, and
the third mode corresponding to the photons of the probe field.
Starting from vacuum, the state at a given time is a fully inseparable three mode
state \cite{mary}.
For certain values of the parameters the state has the same form of Eq.
(\ref{entanstate}), but in general the presence of a third mode
reduces the entanglement between the other two modes.
In the present work we have shown that the collective atomic
recoil lasing from a 3-level atomic BEC can be a more useful
source for the production of the atom-photon entanglement and its application \cite{telebec}

An other potentially interesting situation is when atoms initially occupy both the two
ground states, $|b,0\rangle$ and $|c,0\rangle$, so that the resulting dynamics is
that of a pair of two-level systems coupled by the radiation field.
In fact, if atoms are initially present in $|c,0\rangle$, photons emitted spontaneously by
the transition from $|b,0\rangle$ to $|c,1\rangle$ may drive the other transition between
$|c,0\rangle$ and $|b,-1\rangle$, although detuned by $2\omega_r$ from resonance.
However, if the number of emitted photons is large enough, a fraction of atoms with momentum
$-\hbar\vec q$ will be produced. In the following we discuss in details this effect using
parameters close to those of ref. \cite{Raman:1}.

Taking into account only the four atomic states
$\{|b,0\rangle,|c,0\rangle,|b,1\rangle,|c,-1\rangle\}$ and
treating the bosonic operator as c-numbers, we can derive from
Eq.(\ref{bn})-(\ref{bncn}), the following system of equations:
\begin{eqnarray}
\frac{dS_{1,2}}{dt}&=&-i(\delta\mp\omega_r)S_{1,2}+gAW_{1,2}-\gamma_{1,2}S_{1,2}
\label{MB1}\\
\frac{dW_{1,2}}{dt}&=&-2g(AS_{1,2}^* +{\rm c.c})
\label{MB2}\\
\frac{dA}{dt}&=&gN_b(S_1+S_2)-\kappa A
\label{MB3}
\end{eqnarray}
where $S_1=(b_0c_1^*/N_b)\exp(-i\delta t)$,
$S_2=(b_{-1}c_0^*/N_b)\exp(-i\delta t)$
$W_1=(|b_0|^2-|c_1|^2)/N_b$, $W_2=(|b_{-1}|^2-|c_0|^2)/N_b$,
$A=ae^{-i\delta t}$ and $N_b$ is the number of atoms initially in
the state $|b,0\rangle$. To Eqs.(\ref{MB1}) we have added a
damping term $-\gamma_{1,2}S_{1,2}$ taking into account for the coherence
decay observed experimentally.  Also, we have
added to Eq.(\ref{MB3}) a damping term $-\kappa A$  modelling, in a "mean-field" theory \cite{SR:1},
radiation loss, where $\kappa=cT/2L$ if the
radiation is circulating in a ring cavity (where $T$ is the mirror
transmittivity and $L$ is the cavity length). In the free-space case,
i.e. without optical cavity, $T=1$ and $L$ is of the order of the condensate length.
In the superradiant regime, for $\kappa\gg g\sqrt{N_b}$ and $t\gg
\kappa^{-1}$, we can adiabatically eliminate the radiation field.
Assuming $\delta=\omega_r$ and $\kappa\gg\omega_r$, Eq.(\ref{MB3}) gives
$A\approx (gN_b/\kappa)(S_1+S_2)$, which, when substituted in
Eqs.(\ref{MB1}) and (\ref{MB2}), yields:
\begin{eqnarray}
\frac{dS_1}{dt}&=&-\gamma_1S_1+(G/2)W_1(S_1+S_2)\label{SR:1}\\
\frac{dW_1}{dt}&=&-G\left[2|S_1|^2 +(S_1S_2^*+{\rm c.c})\right]\label{SR:2}\\
\frac{dS_2}{dt}&=&-(\gamma_2+2i\omega_r)S_2+(G/2)W_2(S_1+S_2)\label{SR:3}\\
\frac{dW_2}{dt}&=&-G\left[2|S_2|^2 +(S_1S_2^*+{\rm c.c})\right],\label{SR:4}
\end{eqnarray}
where $G=2g^2N_b/\kappa$ is the superradiant gain. If the number
$N_c$ of atoms initially in the state $|c,0\rangle$ is zero, then
$S_2=0$ and the solution of Eqs.(\ref{SR:1}) and (\ref{SR:2}) yields the
well-known hyperbolic tangent shape for the Superradiant decay of the
fraction of atoms $P_b=|b_0|^2/N_b$ in the initial state $|b,0\rangle$ \cite{LENS}:
\begin{equation}
P_b=1-\frac{1}{2}\left(1-\Gamma\right)
\left\{1+\tanh[G(1-\Gamma)(t-t_D)/2]\right\}
\label{tanh}
\end{equation}
where  $\Gamma=2\gamma_1/G$ and $t_D=[G(1-\Gamma)]^{-1}\ln[N_b(1-\Gamma)]$ is the delay time.
Asymptotically, $P_b$ tends to the stationary value $\Gamma<1$.

In the experiment of ref. \cite{Raman:1}, a cigar-shaped $^{87}Rb$
condensate was illuminated with single laser beam $\pi$ polarized
and detuned by $\Delta_2/(2\pi)=-340$ MHz from the $D_2$ line
transition ($\lambda=780$ nm), between
$|b\rangle=|5^2S_{1/2},F=1,m_F=1\rangle$ and
$|e\rangle=|5^2P_{3/2},F=1,m_F=1\rangle$. After emission of a
photon $\sigma_+$ polarized in the end-fire mode of the
condensate, the atoms return to the ground state
$|c\rangle=|5^2S_{1/2},F=2,m_F=2\rangle$, recoiling at an angle of
$45^0$ with momentum $\vec p=\hbar\vec q$.
The emitted photon is shifted by $-(\Delta_{cb}+\omega_r)$,
where $\Delta_{cb}=(2\pi)6.8$GHz is the shift between the hyperfine ground states and
$\omega_r=\hbar k_2^2/m=(2\pi)7.5$kHz is the recoil frequency.
Normal emission with the
atom back to the same ground state $|b\rangle$ is avoided aligning
the polarization of the laser beam parallel to the main axis of
the condensate. The condensate contained $N_b=10^7$ atoms and had
Thomas-Fermi radii of $R_\parallel=$165 $\mu m$ and $R_\perp=$13.3
$\mu m$, so that $g_1=5\times 10^7/s$ and $g\approx
10^5\sqrt{I}/s$, where $I$ is the laser intensity in $mW/cm^2$.
Assuming $\kappa=c/2R_\parallel\approx 10^{12}/s$, the predicted
superradiant gain is $G/I\approx 2\times 10^5cm^2/(mW s)$. The
measured gain was $G/I\approx 3\times 10^4cm^2/(mW s)$ and the
loss rate was $2\gamma_1=6.2\times 10^4/s$. For $I=7.6 mW/cm^2$,
$\Gamma\approx 0.27$, thus approximately 73\% of atoms were
transferred from the initial state $|b,0\rangle$ to the final
state $|c,1\rangle$, with a momentum ${\vec p}=\hbar{\vec q}$.

Let now consider the effects of having $N_c=\alpha N_b$ atoms in the ground state $|c,0\rangle$,
with initial momentum equal to zero. In fig.\ref{fig2}(a) we show
the results of the numerical integration of Eqs.(\ref{SR:1})-(\ref{SR:4}),
with $G=10\omega_r$, $\gamma_1=\gamma_2=0.3\omega_r$,
and different values of $\alpha=0.1,0.5,1$. We observe that it is
possible to transfer almost 20\% of atoms in the state
$|b,-1\rangle$, moving backward with momentum $\vec p=-\hbar\vec q$.
The fraction of backward atoms is rather small due to the off-resonance
by $2\omega_r$ of the frequency $\omega_1=\omega_2-(\Delta_{cb}+\omega_r)$ of the superradiant field.
Increasing the laser intensity it is possible to make the two populations
of $|b,0\rangle$ and $|b,-1\rangle$ almost equal, if initially $N_b=N_c$. Fig.\ref{fig2}(b) shows the photon
flux per atom, $2\kappa|a|^2/N_b=G|S_1+S_2|^2$, for $\alpha=0.1,0.5,1$. The radiation peak reduces
increasing $\alpha$, because the absorbtion from the second transition becomes more important.

In conclusion, we presented a quantum theory describing the experimentally observed
superradiant Raman scattering from a Bose-Einstein condensate
driven by a single off-resonant laser beam. We showed that
collective atomic recoil lasing (CARL) from 3-level atoms in a
$\Lambda$-configuration, realized using two hyperfine levels of the
ground state, produces Raman amplification of matter waves.
In particular, when atoms are initially in one of the two lower states,
a pure two-level system is realized between atoms with
different internal states and different momentum, and entangled
atom-photon pairs are generated. In this case, the system behaves as a
non-degenerate optical parametric amplifier.
When the atoms are initially in both the hyperfine levels of the ground state, photons emitted
superradiantly by atoms in the first two-level system can be absorbed by atoms in the second two-level system,
generating a condensate recoiling in the backward direction.
We observe that in this case it should be possible to measure experimentally any
eventual difference between decoherence rates for atoms recoiling in opposite directions.
In fact, a recent experiment \cite{LENS} gave evidence of a phase-diffusion contribution to atomic decoherence
depending on the detuning from the two-photon Bragg resonance condition.
In the present case, superradiant photons resonantly emitted in one transition do not satisfy the resonance
condition for the other transition. Hence, it should be possible to evaluate the phase-diffusion contribution to
decoherence measuring the final steady-state fraction of atoms in the two recoiling condensates.

\newpage

\begin{figure}
\begin{center}
\includegraphics[width=10cm]{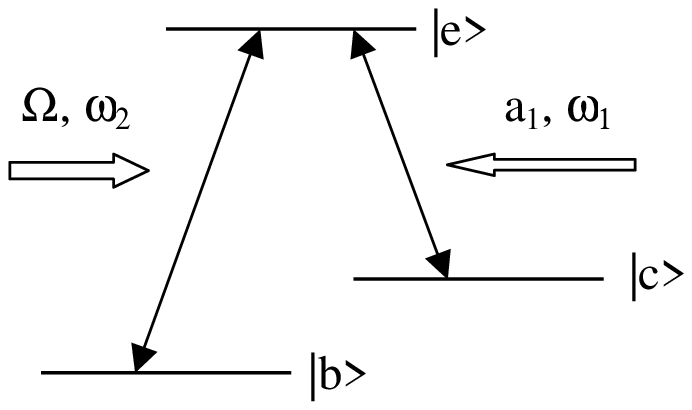}
\end{center}
\vskip 0.2cm \caption{Three-level $\Lambda$-shaped atoms coupled
to a quantized probe laser $a_1$ and a classical coupling laser $\Omega$
with frequency $\omega_1$ and $\omega_2$, respectively.} \label{fig1}
\end{figure}

\begin{figure}
\begin{center}
\includegraphics[width=10cm]{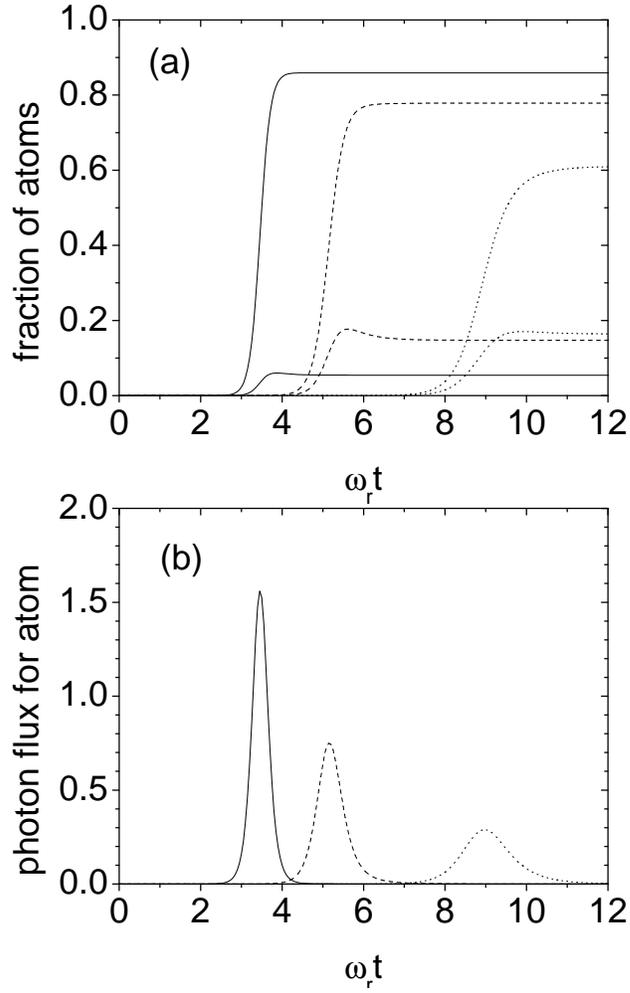}
\end{center}
\vskip 0.2cm \caption{(a): Fraction of atoms in $|c,1\rangle$ (upper
curves) and in $|b,-1\rangle$ (lower curve) vs. $\omega_r t$, for $G=10\omega_r$,
$\gamma_1=\gamma_2=0.3\omega_r$ and $\alpha=N_c/N_b=0.1$
(continuous lines), $\alpha=0.5$ (dashed lines) and $\alpha=1$
(dotted lines). (b): photon flux for atom, $2\kappa|a|^2/N_b=G|S_1+S_2|^2$,
for $\alpha=0.1$ (continuous line), $\alpha=0.5$ (dashed line) and $\alpha=1$ (dotted line).}
\label{fig2}
\end{figure}

\end{document}